# QCD QED Potentials, Quantum Field Theoretical Generalization of Yukawa Potential

Eue-Jin Jeong
University of Texas at Austin
Austin TX 78741 USA

## Abstract

Despite the success of quantum field theories, the origin of the mass of elementary particles persists. The renormalization program is an essential part of the calculation of the scattering amplitudes, where the infinities of the calculated masses of the elementary particles are subtracted for the progressive calculation of the higher-order perturbative terms. The mathematical structure of the mass term from quantum field theories expressed in the form of infinities suggests that there exists a finite dynamical mass in the limit when the input mass parameter approaches zero. The Lagrangian recovers symmetry at the same time as the input mass becomes zero, whereas the self-energy diagrams acquire a finite dynamical mass of the quantum fields in the 4-dimensional space when the dimensional regularization method of renormalization is utilized. The complex forms of the QCD and QED interaction potentials are obtained by replacing the fixed mass and coupling constants in the Yukawa potential with the scale-dependent running coupling constant and the corresponding dynamical mass. The derived QCD potential predicts quark confinement and deconfinement, and the QED potential derived by the same method predicts the sharply rising delta function potential near the contact distance between the electron and positron.

## 1. Introduction

The standard Glashow-Weinberg-Salem (1) model of electroweak interaction has been highly successful in predicting the interactions of high-energy elementary particles. The discovery (2) of the W and Z gauge bosons, and finally the discovery of the Higgs boson at CERN in 2012 (3), proved that the standard model is a mathematically correct theory describing the interactions of elementary particles.

However, a consistent interaction potential model has not been proposed for QCD and QED. We investigated the structures of the self-energy diagrams of the elementary particles to study the relationship between the mass and coupling constant in quantum field theories and apply them to construct the interaction potential model. By using the dimensional regularization method for the renormalization of quantum field theories, a finite indeterminate mathematical form of the dynamical mass of the fields is obtained in the limit of the input mass term in the Lagrangian approaches zero in the dimensional regularization method. In this process, the symmetry of the original Lagrangian is restored, whereas a finite mass appears in the self-energy-loop diagrams. The renormalization group equation (4) resolves the problem of arbitrariness of the renormalization prescription. The dynamic mass generation mechanism is presented within the framework of the dimensional regularization method developed by G. 't Hooft and M. Veltman (5).

## 2. Dynamical Mass from the Massless Quantum Field Theory

(1) $\lambda\varphi^4$ theory

The mathematical structure of the one loop self-energy diagram in $\lambda\varphi^4$ theory is represented by

$$\text{One loop} = \frac{m_0^2 \lambda}{16\pi^2}\left[\frac{1}{n-4} + \frac{1}{2}\psi(2) - \frac{1}{2}\ln\frac{m_0^2}{4\pi\mu^2} + O(n-4)\right], \quad [1]$$

where $\psi(2)$ is a constant given in general,

$$\psi(n+1) = 1 + \frac{1}{2} + \ldots + \frac{1}{n} - \gamma \quad (\gamma = 0.5772\ldots)$$

, and $\mu$ is an arbitrary constant with mass dimensions. Renormalization for nonzero bare mass $m_0$ is necessary because the first term is divergent in the $n \to 4$ limit. However, in the zero bare mass limit $m_0 \to 0$, the term is not infinite but becomes undetermined. We introduce a constant $C_s$, and the one-loop diagram becomes

$$\lim_{\substack{m_0 \to 0 \\ n \to 4}}\left[m_0^2 \frac{\lambda}{16\pi^2}\left[\frac{1}{n-4} + \frac{1}{2}\psi(2) - \frac{1}{2}\ln\frac{m_0^2}{4\pi\mu^2} + O(n-4)\right]\right] = \lim_{\substack{m_0 \to 0 \\ n \to 4}} \frac{\lambda}{16\pi^2}\frac{m_0^2}{n-4} \equiv \frac{\lambda}{16\pi^2}C_s,$$

[2], where

$$C_s = \lim_{\substack{m_0 \to 0 \\ n \to 4}} \frac{m_0^2}{n-4}.$$

As a result of this operation, we have an analytical mass that is not infinity but simply undetermined. Therefore, the massless $\lambda\varphi^4$ scalar field theory begins to have a mass from a one-loop self-energy diagram. Recalling that the $\lambda\varphi^4$ massless scalar field theory is the simplest case of supersymmetric theories, it provides us with a clue to a possible mass-generation mechanism for supersymmetric particles.

The fact that the explicit mass parameter in the Lagrangian does not represent the real mass of the field and its sole purpose is to provide a reference from which the real mass can be determined experimentally has already suggested that the mass can be generated by dynamical interactions of the interacting fields. In the case of QCD and QED, the self-energy was calculated without explicit mass parameters in the Lagrangian.

(2) QED

The self-energy diagram of the electron in QED (7) without the mass parameter in the Lagrangian is given by

$$\Sigma(p) = \frac{2}{n-4}\frac{e^2}{16\pi^2}P - \frac{e^2}{8\pi^2}\left[\frac{1}{2}P(1+\gamma) + \int_0^1 dx P(1-x)\ln\frac{p^2 x(1-x)}{4\pi\mu^2}\right] \quad [3]$$

where P represents the energy-momentum tensor of the electronic quantum field. Because self-energy is defined by the energy when the particle is in a rest state, the mass of the electron is given by

$$M_e = \frac{e^2}{8\pi^2} C_e \quad [4]$$

Where $C_e = \lim_{n \to 4}\frac{p \to 0}{}[\frac{\det P}{n-4}]$ and the higher-order small correction terms can be included in the constant factor $C_e$, without the loss of generality.

The vacuum polarization diagram of the photon (7) is given by

$$\Pi_{\mu\nu}(p) = \frac{e^2}{2\pi^2}(P_\mu P_\nu - \delta_{\mu\nu}P^2)[\frac{1}{3(n-4)} - \frac{1}{6}\gamma - \int_0^1 dx x(1-x)\ln\frac{p^2 x(1-x)}{2\pi\mu^2}] + O(n-4) \quad [5]$$

The dynamical mass of the photon is now given by $M_\gamma^2 = \frac{e^2}{6\pi^2} C_\gamma$
where the photon mass constant $C_\gamma = \lim_{n \to 4}\frac{p \to 0}{}[\frac{\det(P_\mu P_\nu - \delta_{\mu\nu}P^2)}{n-4}]$. Although it is generally known that photons do not carry mass, the gauge invariance of the Lagrangian suggests that they manifest mass in relation to the distance of their interactions with the charged particles. In fact, the self-energies of photons and gluons manifest themselves as masses.

(3) QCD

Using the same procedure for QCD, the dynamical mass for quarks is given by

$$M_f = C_3 \frac{g^2}{8\pi^2} C_f \qquad C_f = \lim_{n \to 4}\frac{p \to 0}{}[\frac{\det P}{n-4}] \quad [6]$$

and

$$M_{Y.M}^2 = (\frac{5}{3}C_1 - \frac{4}{3}C_2)\frac{g^2}{8\pi^2} C_{Y.M} \qquad C_{Y.M} = \lim_{n \to 4}\frac{p \to 0}{}[\frac{\det(P_\mu P_\nu - \delta_{\mu\nu}P^2)}{n-4}] \quad [7]$$

for the self-energy of the gluon from Yang Mill fields.

## 3. Self-Energy and Coupling Constant in the Quantum Fields

It is well known that the electron mass is related to the electrostatic self-energy in classical electrodynamics, where the radius $r_0$ of the electron is defined by

$$m_e = \frac{e^2}{r_0}. \qquad [8]$$

In fact, the relationship between the mass and the corresponding charge of a particle is a universal feature beyond classical electrodynamics. The quantum field theoretical dynamic mass is directly related to the corresponding coupling constants by the following relations, as shown in the above examples:

$$M_s^2 = \frac{\lambda}{16\pi^2} C_s \qquad C_s = \lim_{n \to 4}^{m_o \to 0}\left[\frac{m_o^2}{n-4}\right] \qquad [2]$$

$$M_e = \frac{e^2}{8\pi^2} C_e \qquad C_e = \lim_{n \to 4}^{p \to 0}\left[\frac{\det P}{n-4}\right] \qquad [4]$$

$$M_\gamma^2 = \frac{e^2}{6\pi^2} C_\gamma \qquad C_\gamma = \lim_{n \to 4}^{p \to 0}\left[\frac{\det(P_\mu P_\nu - \delta_{\mu\nu} P^2)}{n-4}\right] \qquad [5]$$

$$M_f = C_3 \frac{g^2}{8\pi^2} C_f \qquad C_f = \lim_{n \to 4}^{p \to 0}\left[\frac{\det P}{n-4}\right] \qquad [6]$$

$$M_{Y.M}^2 = \left(\frac{5}{3}C_1 - \frac{4}{3}C_2\right)\frac{g^2}{8\pi^2} C_{Y.M} \qquad C_{Y.M} = \lim_{n \to 4}^{p \to 0}\left[\frac{\det(P_\mu P_\nu - \delta_{\mu\nu} P^2)}{n-4}\right] \qquad [7]$$

where $C_1$, $C_2$, and $C_3$ are constants determined by the group structure of the non-Abelian gauge theory, and the sub-indices s, e, $\gamma$, f, and Y.M indicate the scalar, electron, photon, fermion, and Yang-Mills fields, respectively. In the four-dimensional space, all constants, including the higher-order correction terms for the self-energies, become undetermined in the limit of the momentum, and the input mass becomes zero. These relations between the mass and coupling constant suggest a significant variation in the mass due to the running coupling constant that depends on the scale.

## 4. QCD and QED Potentials by Generalizing Yukawa Potential

In 1935, Yukawa (6) introduced the nuclear potential, which has been proven to be highly successful in addressing many of the diverse nuclear interactions. The major property of Yukawa's potential is the introduction of the mass of the pion as the interaction-mediating particle, which applies for strong nuclear force at short distances. The coupling constant and mass of the pion in Yukawa's nuclear potential are independent fixed parameters regardless of the mutual interaction distance.

$$V_{Yukawa}(r) = -g^2 \frac{e^{-\alpha m r}}{r} \qquad [9]$$

where $g$ is the coupling constant, $m$ is the mass of the intermediate particle, $r$ is the radial distance between particles, and $\alpha$ is a scaling constant.

(1) QCD

Because we have established the dependence between the scale-dependent coupling constant and the self-energy of the quantum fields, we propose constructing a new generalized Yukawa potential by replacing the fixed mass and coupling constant with those that depend on the running coupling constant [2]–[7]. The generalized Yukawa potential with the variable-scale-dependent running coupling constant and the corresponding self-energy is given by

$$V(r) = g^2(\mu) \frac{e^{-\alpha m(\mu) r}}{r} \qquad [10]$$

Where $g^2(\mu)$ is the running coupling constant and $m(\mu)$ is the scale-dependent self-energy (mass) of the interaction-mediating field in the QFT. For example, photon mass is zero in macroscopic scale and Yukawa potential with zero mass interaction mediating particle takes the form of Coulomb potential. This property of Yukawa potential indicates that the fundamental mathematical structure of Yukawa potential is much more general than typically known as nuclear potential. It has been shown that the coupling constant and mass of the fields depends on the scale in quantum field theory. Therefore, using the property of the generality of Yukawa potential, it must be possible to derive a detailed form of QCD and QED potentials that are effective in sub-hadronic scale by utilizing the mathematical form of the scale dependent coupling constant and the mass of the interaction mediating particles in quantum field theories. It is difficult to predict what will be the outcome at this point. We expect that there could be something unique that can explain the physical phenomena that have not been possible to decipher in the sub-hadronic realm especially the quark confinement problem. The scale-dependent running coupling constant from QCD is given by

$$g^2(\mu) = \frac{g_0^2}{1 + \frac{g_0^2}{8\pi^2} \ln \frac{\mu}{\mu_0}} \qquad [11]$$

which was developed by D. Gross, F. Wilzeck, and H. D. Politzer (8) (9) and using the scale-dependent self-energy of Yang-Mill field [7], the QCD potential is given by

$$V_{qcd}(r) = \frac{g_0^2}{1 + \frac{g_0^2}{8\pi^2} \ln \frac{\mu}{\mu_0}} \frac{1}{r} \exp\left(-\left(\frac{\frac{\alpha}{8\pi^2} g_0^2 C_g C_k}{1 + \frac{g_0^2}{8\pi^2} \ln \frac{\mu}{\mu_0}}\right)^{1/2} r\right) \qquad [12]$$

where $C_g$ is the gluon mass constant, which is given by $C_g = 5.8 \times 10^{-103} g^2$, and $C_\kappa$ is a group structure constant of an order of magnitude 1. However, the potential in the form [12] is impractical because of the parameter $\mu$, which depends on the input momentum scale. To translate the parameter $\mu$ into distance $r$, we hypothesize that there is a mathematical relationship between $\mu$ and $r$ governed by

$$\mu = \lambda \exp\left(\frac{\varrho}{r^2}\right) \quad \lambda, \varrho > 0 \qquad [13]$$

where $\lambda$ and $\rho$ are the adjustable constants. The relation [13] does not violate the quantum mechanical uncertainty because the larger input momentum $\mu$ results in a smaller distance $r$ owing to the quantum uncertainty principle,

$$\Delta x \Delta p \geq \hbar/2 \qquad [14]$$

In fact, the mathematical relation [13] is the only possible choice to obtain $1/r$ dependent Coulomb potential at large distances and the QED potential at sub-hadronic distances that confirms the phenomenological quarkonia spectroscopy results (19).

After the transformation of $\mu$ by the relation [13], the QCD potential [12] is given by:

$$V_{qcd}(r) = \frac{1}{\frac{A}{r^2}-B}\frac{1}{r}\exp\left(-\left(\frac{\frac{\alpha}{8\pi^2}C_k C_g}{\frac{A}{r^2}-B}\right)^{1/2} r\right) \quad [15]$$

where $A=\frac{g_0^2 \rho}{8\pi^2}$, $B=\frac{\ln(\frac{\mu_0}{\lambda})}{8\pi^2}-\frac{1}{g_0^2}$, and $C_g$ is the gluon mass constant which is given by $C_g = 5.8 \times 10^{-103} g^2$, and $C_\kappa$ is a group structure constant of the order of magnitude 1.

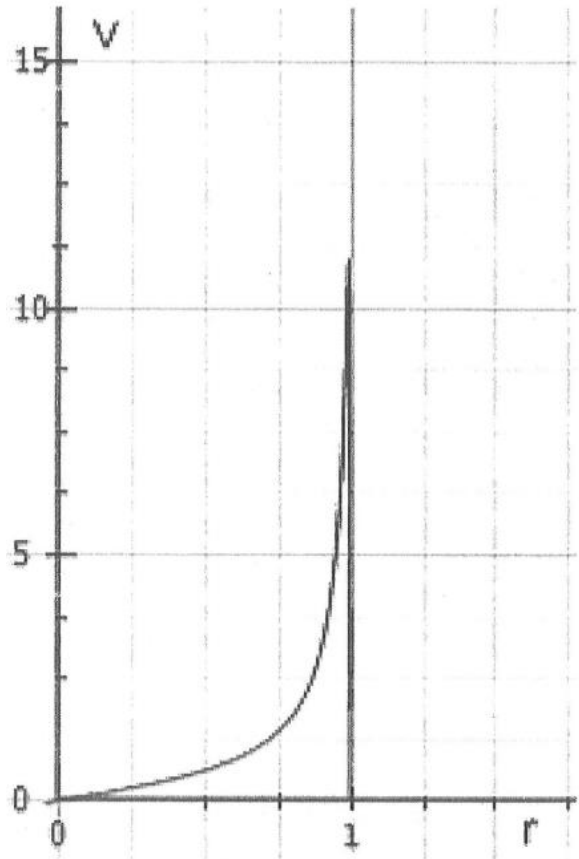

**Fig. 1 QCD Potential Diagram**

At $r = r_{cq} = \sqrt{A/B}$, the QCD potential [15] becomes zero due to the negative infinite exponential factor and becomes imaginary as $r$ increases further. To visualize the general structure of the potential, for instance, for $A=B=1$ and $\frac{\alpha}{8\pi^2}C_\kappa C_g = 0.05$, the QCD potential has the form depicted in the diagram in Fig. 1, which shows the initial confinement and deconfinement by the sharply dropping potential after reaching the maximum and the decay phase as the potential becomes imaginary below zero level. In quantum mechanics, imaginary potential is known to violate the conservation of the probability of finding quantum particles. The loss of probability beyond the outer radius of the hadronic boundary is consistent with the decay of the quarks and also with the spontaneous evaporation (11) of the black hole at its surface, assuming that the black hole is fundamentally a neutron star with an extreme density of quark-gluon plasma.

For small $r$ and $\alpha = 1$, the QCD potential [15] becomes

$$V_{qcd}(r) = \frac{r}{A}(1 - \frac{B}{A}r^2 + ......)  \quad [16]$$

The quark potential [15] shown in Fig. 1 has the following features.

1. linear potential at small distance $r$
2. confinement within the hadronic boundary
3. deconfinement beyond the critical distance of the hadronic boundary as the potential drops to zero
4. decay (disappearance) of quark matter as the potential becomes imaginary as the relative distance increases beyond the zero-potential level
5. no singularity throughout the relative distances

(2) QED

By applying the same mathematical procedure using the running coupling constant for the QED

$$e^2(Q) = \frac{e^2}{1 - \frac{e^2}{6\pi^2} \ln \frac{Q}{m_e}} \quad [17]$$

Where $Q$ is the input momentum scale given by $Q = \lambda \exp(\frac{\rho}{r^2})$ for the transformation into the length parameter $r$, and $\lambda$, $\varrho$ are adjustable constants that have the same form as in the case of QCD [13], the QED potential is given by:

$$V_{qed}(r) = \frac{-1}{C - \frac{D}{r^2}} \frac{1}{r} \exp(-(\frac{\frac{\alpha}{6\pi^2} C_\gamma}{C - \frac{D}{r^2}})^{1/2} r) \quad [18]$$

where, $C = \frac{1}{e^2} + \frac{\ln(\frac{m_e}{\lambda})}{6\pi^2}$, $D = \frac{e^2\rho}{6\pi^2}$ and $C_\gamma$ is the photon mass constant [5] with the upper limit of the photon rest mass $3 \times 10^{-53} g$ (10).

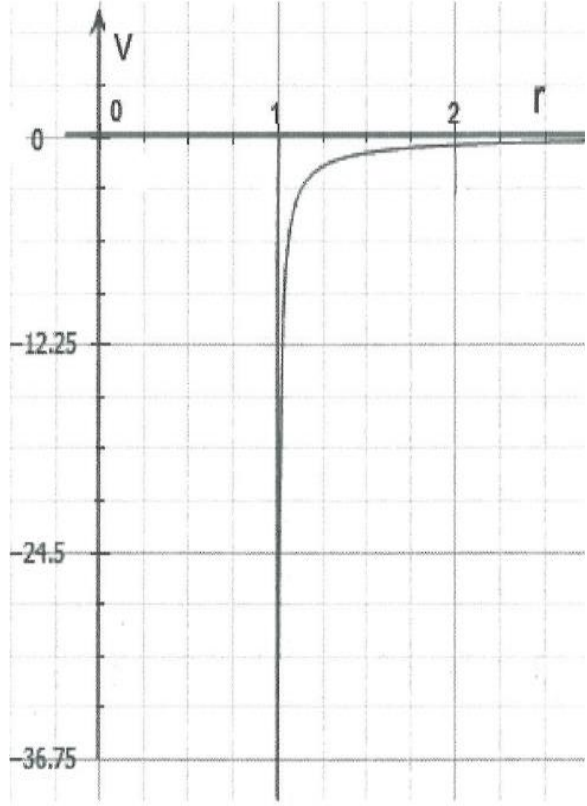

**Fig. 2 QED Potential Diagram**

To visualize the detailed structure of the QED potential, for instance, for $C=1$, $D=1$, and $\frac{\alpha}{6\pi^2}C_\gamma = 0.01$, the potential is depicted in Fig. 2 which shows the sharply rising core potential at the contact boundary of the electron and positron at $r = \sqrt{D/C} = 1$ as they approach close together. The potential becomes imaginary as both particles come close past the contact distance and reach a level above zero. This behavior of the QED potential is consistent with the electron-positron pair annihilation as they approach sufficiently close together. It should be noted that the sharply rising core potential has been employed for the calculation of the "Lamb shift" (12) in the form of the $\delta(r)$ function in the phenomenological model. The parameters $C$ and $D$ determine the contact radius of the electron-positron pair, and $\frac{\alpha}{6\pi^2}C_\gamma$ determines the depth of the QED potential. By adding the two potentials [15] and [18] at typical low-energy hadronic bound states, we obtain

$$V(r) = \frac{-1}{Cr} + \frac{r}{A} \quad\quad [19]$$

This result confirms the previously reported non-relativistic phenomenological quark potential, which was in good agreement with the experimental results in heavy quarkonia spectroscopy (13).

The QCD potential [15] in Fig. 1 shows a small yet finite probability of finding fractional charges beyond the critical distance, which supports the results reported by researchers (14), even though the quark itself has not been isolated. In the case of the QED potential with electron-positron annihilation, the interaction of the electron with the antimatter positron is considered the key to the loss of the quantum probability of the electron at close distances. However, in the case of QCD, the quark's loss of quantum probability beyond the distance of the hadronic boundary despite the apparent absence of nearby anti-quark matter is considered a mystery, although the derived QCD potential confirms the experimental data and is consistent with the predicted spontaneous decay of the black hole (11), which is essentially a large-scale matter state of the quark-gluon plasma.

## Conclusion

We presented a uniform mathematical procedure to transform perturbative quantum field theories into unified interaction potential model for both QCD and QED by utilizing the running coupling constant derived from the renormalization group equations within the framework of the known Yukawa nuclear potential model. Both the QCD and QED potentials show sharply reversing curvature of the peak potential at the critical distance without losing continuity and these two different types of potentials confirm the experimental data at all ranges including the loss of probability by becoming imaginary beyond the critical distances.

## Acknowledgement

I would like to thank Dr. Martinus Veltman for his memorable lectures on the dimensional regularization method for renormalization.

## References

(1) S. L. Glashow, Nucl. Phys. 22 (1961) 579, A. Salam and J. C. Ward, Phys. Lett. 13 (1964) 1681, S. Weinberg, Phys. Rev. Letter. 19 (1967) 1264, A. Salam, in: Proc. 8$^{th}$ Nobel Symp., 2$^{nd}$ Ed., ed.,N. Svartholm (Almquist and Wikasells, Stockholm, 1978) p. 367.

(2) UA-1 Collab., G. Arnison et al., Phys. Lett. 122B (1983) 103; 126B (1983) 398; 129B (1983) 273; UA-2 Collab., G. Banner et al., Phys. Lett.122B (1983) 476; 129B (1983) 130

(3) ATLAS Collaboration, Physics Letters B Volume 716, Issue 1, 17 (2012)

(4) Gell-Mann and Low, Phys. Rev. 95 (1954) 1300; C. Callan, Phys. Rev. D5 (1973) 3202; K. Symanzik, Comm. Math. Phys. 23 (1971) 49; 't Hooft, Nucl. Phys. B61 (1973) 455; S. Weinberg, Phys. Rev. D8 (1973) 3497; S. Coleman, Lecture Erice summer school

(1971); K. Wilson, Revs. Modern Phys. 47 (1975) 773; E.C.G. Stueckelberg and A. Peterman, Helv. Phys. Acta 26 (1953) 499; K. Wilson, Phys. Rev. D3 (1971) 1818.

(5) G. 't Hooft and M. Veltman, Nuclear Physics B44 (1972) 189- 213

(6) H. Yukawa. (1935). "On the interaction of elementary particles". Proc. Phys. Math. Soc. Japan. **17**: 48.

(7) P. Ramond, Field Theory, A Modern Primer Benjamin Cummings pub./com. Inc. (1980).

(8) M. Gell-Mann and F. Low, Phys. Rev. 95 (1954) 1300; E. Stueckelberg and A. Petermann, Helv. Phys. Acta 5 (1953) 499; N. N. Bogoliubov and D. V. Shirkov, Introduction to the theory of Quantized Fields ( Interscience, New York, 1959); K. Wilson, Phys. Rev. D3 (1971) 1818; K. Wilson and J. Kogut, Phys. Reports 12 (1974) 75; K. Symanzik, Lett. Nuovo Cimento 6 (1973) 77; D. Gross and F. Wilczek, Phys. Rev. Letters 26 (1973) 1343; H. D. Politzer, Phys. Rev. Letters 26 (1973) 1117.

(9) D. Gross and F. Wilczek, Phys. Rev. Lett. 26 (1973) 1343 H. D. Politzer, Phys. Rev. Lett. 26 (1973) 1346

(10) A. Barnes and J. Scargle, Phys. Rev. Lett. 35 (1975) 1117

(11) S. W. Hawking, S. W., "Black hole explosions?" Nature. 248 (1974) 30 and "Particle creation by black holes" Commun. Math. Phys. 43 (1975) 199; Christian Corda, Classical and Quantum Gravity. 32 (2015) 195007
(12) Bjorken & Drell, Relativistic Quantum Mechanics, McGraw-Hill Inc. (1964).

(13) K. J. Miller and M. G. Olsson, Phys. Rev. 25D (1982) 2383; E. Eichten, K, Gottfried, T. Kinoshita, J. Kogut, K. D. Lane and T. M. Yan, Phys. Rev. Lett. 34 (1975) 369; S. Jacobs and M. G. Olsson, Phys. Lett. 133B (1983) 111; E. Eichten, K. Gottfried, T. Kinoshita, K. D. Lane and T. –M. Yan, Phys. Rev. 21D (1980) 203; V. A. Novikov et al., Phys. Reports 41C (1978) 16.; A Manohar, H Georgi- Nuclear Physics B. 234 (1984) 189.; M Abu-Shady, SY Ezz-Alarab- Few-Body Systems. 3 (2019) 186; R Bonnaz, B Silvestre-Brac, C Gignoux- Eur. Phys. J. A. 13 (2002) 363

(14) G. S. LaRue, W. M. Fairbank and A. F. Hebard Phys. Rev. Lett. 38 (1977) 1011; G. S. LaRue, W. M. Fairbank, and J. D. Phillips, Phys. Rev. Lett. 42 (1979) 142; L. J. Schaad and B. A. Hess, Jr., J. P. Wikswo, Jr., W. M. Fairbank, Phys. Rev. 23A (1981) 1600; GL Shaw, R Slansky, Phys. Rev. Lett. 50 (1983) 1959; Yuval Gefen and David J. Thouless. Phys. Rev. B 47, (1993) 10423; Eun-Ah Kim, Michael Lawler, Smitha Vishveshwara, and Eduardo Fradkin Phys. Rev. Lett. 95, (2005) 176402; Eun-Ah Kim, Michael J. Lawler, Smitha Vishveshwara, and Eduardo Fradkin Phys. Rev. B 74 (2006) 155324; XL Qi, Xiao-Liang Qi, Taylor L. Hughes & Shou-Cheng Zhang, Nature Physics 4, (2008) 273; Jie Zhou, Wuhong Zhang, and Lixiang Chen Appl. Phys. Lett. 108 (2016) 111108